\documentclass[aps,twocolumn, amsmath,amssymb, superscriptaddress]{revtex4-1}

\usepackage{hyperref}

\usepackage[utf8]{inputenc}
\usepackage{lmodern}
\usepackage[version=3]{mhchem}
\usepackage{amsmath,amssymb,amstext,amsfonts}
\usepackage{natbib}
\usepackage{psfrag,graphicx}
\usepackage[]{units}
\usepackage{upgreek}
\usepackage{textcomp} 
\usepackage{color}
\usepackage{float}
\usepackage{subfigure}
\usepackage{array} % Matrizen in mathematischen Formeln
\usepackage{epstopdf}
\usepackage[ngerman, num]{isodate}

\begin{document}
% ------
% Maketitle metadata
\title{3D Print of Polymer Bonded Rare-Earth Magnets, and 3D Magnetic Field Scanning With an End-User 3D Printer}

\author{C. Huber}
\thanks{Correspondence to: \href{mailto:christian.huber@tuwien.ac.at}{christian.huber@tuwien.ac.at}}
\affiliation{Institute of Solid State Physics, Vienna University of Technology, 1040 Vienna, Austria}
\affiliation{Christian Doppler Laboratory for Advanced Magnetic Sensing and Materials, 1040 Vienna, Austria}

\author{C. Abert}
\affiliation{Institute of Solid State Physics, Vienna University of Technology, 1040 Vienna, Austria}
\affiliation{Christian Doppler Laboratory for Advanced Magnetic Sensing and Materials, 1040 Vienna, Austria}

\author{F. Bruckner}
\affiliation{Institute of Solid State Physics, Vienna University of Technology, 1040 Vienna, Austria}
\affiliation{Christian Doppler Laboratory for Advanced Magnetic Sensing and Materials, 1040 Vienna, Austria}

\author{M. Groenefeld}
\affiliation{Magnetfabrik Bonn GmbH, 53119 Bonn, Germany}

\author{O. Muthsam}
\affiliation{Institute of Solid State Physics, Vienna University of Technology, 1040 Vienna, Austria}
\affiliation{Christian Doppler Laboratory for Advanced Magnetic Sensing and Materials, 1040 Vienna, Austria}

\author{S. Schuschnigg}
\affiliation{Department of Polymer Engineering and Science, Montanuniversitaet Leoben, Leoben, Austria}

\author{K. Sirak}
\affiliation{Institute of Solid State Physics, Vienna University of Technology, 1040 Vienna, Austria}
\affiliation{Christian Doppler Laboratory for Advanced Magnetic Sensing and Materials, 1040 Vienna, Austria}

\author{R. Thanhoffer}
\affiliation{Institute of Solid State Physics, Vienna University of Technology, 1040 Vienna, Austria}
\affiliation{Christian Doppler Laboratory for Advanced Magnetic Sensing and Materials, 1040 Vienna, Austria}

\author{I. Teliban}
\affiliation{Magnetfabrik Bonn GmbH, 53119 Bonn, Germany}

\author{C. Vogler}
\affiliation{Institute of Solid State Physics, Vienna University of Technology, 1040 Vienna, Austria}

\author{R. Windl}
\affiliation{Institute of Solid State Physics, Vienna University of Technology, 1040 Vienna, Austria}
\affiliation{Christian Doppler Laboratory for Advanced Magnetic Sensing and Materials, 1040 Vienna, Austria}

\author{D. Suess}
\affiliation{Institute of Solid State Physics, Vienna University of Technology, 1040 Vienna, Austria}
\affiliation{Christian Doppler Laboratory for Advanced Magnetic Sensing and Materials, 1040 Vienna, Austria}

\date{20.05.2016}
\date{\today}

$ $\newline
%%%%%%%%%%%%%%%%%%%%%%%%

%\thispagestyle{empty}

\begin{abstract}
3D print is a recently developed technique, for single-unit production, and for structures that have been impossible to build previously. The current work presents a method to 3D print polymer bonded isotropic hard magnets with a low-cost, end-user 3D printer. Commercially available isotropic NdFeB powder inside a PA11 matrix is characterized, and prepared for the printing process. An example of a printed magnet with a complex shape that was designed to generate a specific stray field is presented, and compared with finite element simulation solving the macroscopic Maxwell equations. For magnetic characterization, and comparing 3D printed structures with injection molded parts, hysteresis measurements are performed. To measure the stray field outside the magnet, the printer is upgraded to a 3D magnetic flux density measurement system. To skip an elaborate adjusting of the sensor, a simulation is used to calibrate the angles, sensitivity, and the offset of the sensor.  With this setup a measurement resolution of 0.05\,mm along the z-axes is achievable. The effectiveness of our novel calibration method is shown. 

With our setup we are able to print polymer bonded magnetic systems with the freedom of having a specific complex shape with locally tailored magnetic properties. The 3D scanning setup is easy to mount, and with our calibration method we are able to get accurate measuring results of the stray field.

\end{abstract}

\maketitle

\section{Introduction}
Polymer bonded magnets have opened a new world of application opportunities in the sensor and electric drive technology \cite{ehrenstein_paper, ibb}. By modification of thermosoftening plastic with hard magnetic filler particles it is possible to manufacture polymer bonded permanent magnets. As hard magnetic particles, ferrite (e.g. Sr, Ba) as well as rare-earth materials (e.g. NdFeB) with a volume filler content between 45 - 65\,\%  are inserted. First, the polymer and the magnetic particles are compounded in a twin-screw extruder. Then, the obtained compound can be further processed with injection molding and extrusion, respectively \cite{diss_drummer}.

NdFeB powder for polymer bonded magnets can be produces by a melt spinning process. This produces ribbons or flakes with a size about 200\,\textmu m \cite{develop_process, coated_ndfeb}. Inert gas atomization processes produce spherical powder with a particle size of approximately 45\,\textmu m \cite{recent_devel}, which is usually preferred in injection molding processes to achieve better rheological behavior. Whenever a high maximum energy product $(BH)_{max}$ of the bonded magnets is not the most significant characteristic value, magnetically isotropic powder is preferred because it comes with lower assembling costs and more flexibility.

Conventional magnets are mainly produced by sintering, and hence limited in the complexity of their shapes. Due to procedural advantages of plastic technologies, polymer bonded hard magnets enable the manufacturing of complex shapes and features by design flexibility regarding shape and magnetizing structure. However, $(BH)_{max}$ of these magnets is barely half of sintered one, as well as these technologies are only affordable, and economically reasonable for mass production of permanent magnets \cite{recent_devel}. Currently, no single-unit production technologies are available to produce magnets with complex structures. For that reason, we present in this work a novel cost and time effective manufacturing process for polymer bonded rare-earth magnets with an arbitrary shape.

Since 3D printers are nowadays affordable for end-users, a boom of new possibilities has been triggered. 3D print technology is a fast growing field for single-unit production, and it allows to produce structures that have been historically difficult or impossible to build, like holes that change direction, unrealistic overhangs, or square interior cavities.

%Furthermore we present a novel calibration method for a 3D Hall sensor. To measure the stray field of our printed magnets a 3D Hall sensor is attached to the printer head. Therefore, the 3D printer was upgraded to a magnetic flux density scanner. To skip an elaborate adjusting of the sensor, a simulation was used to calibrate the mounting angles and offset of the sensor and the electrical sensitivity. 

%With our setup we are able to print affordable isotropic polymer bonded NdFeB magnetic systems with unprecedented shapes. The scanning setup is easy to mount, and with our calibration method we are able to get accurate measuring results.

\section{Material}
%\label{sec:material}
For injection molding of highly filled plastics, the fluidity of the matrix material is essential. This is due to rising filler content which increases the viscosity of the compound leading to filling problems in the cavity. For this reason, the matrix material should be of a high flowable material as well as good mechanical properties. Polyamide has a good combination of these properties, and is therefore suited for the processing of highly filled plastics.  Especially polyamides such as PA6, PA11, and PA12 are commercially relevant \cite{diss_drummer, abs}.

In our case a prefabricated compound (Neofer\,\textregistered$ $ 25/60p) from Magnetfabrik Bonn GmbH is used. Neofer\,\textregistered$ $ 25/60p is a compound of 
 NdFeB grains with uniaxial magnetocrystalline anisotropy inside a PA11 matrix. The orientation of the NdFeB grains is random leading to isotropic magnetic properties of the bulk magnet. The powder is produced by employing an atomization process followed heat treatment.
 
\begin{figure}
	\centering
	\includegraphics[width=\linewidth]{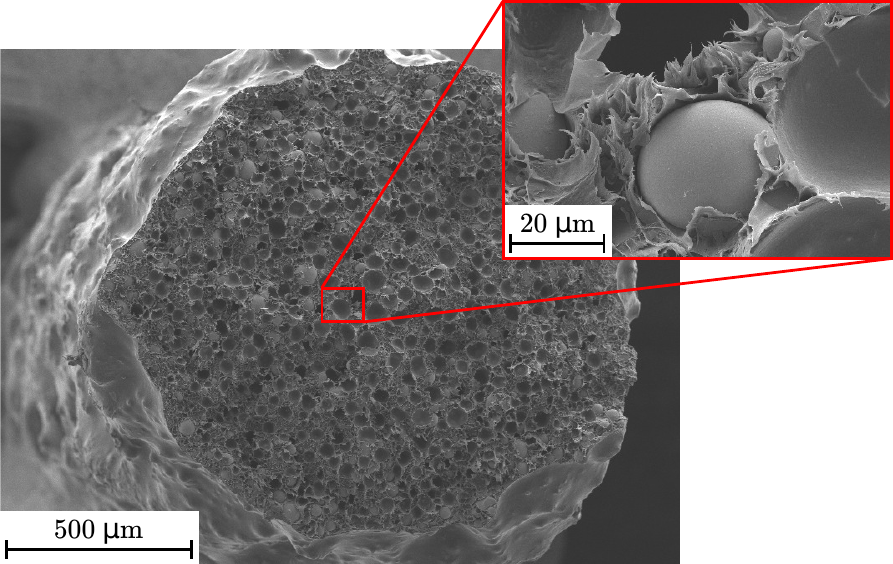}
	\caption{SEM micrograph of the Neofer\,\textregistered$ $ 25/60p filament, NdFeB spheres inside the PA11 matrix.}
	\label{fig:neofer_sem}
\end{figure}

 Thermogravimetric analysis (TGA) measurement yields a filler content of 90\,wt\%. The magnetic datasheet values of the material are listed in Tab.\ref{tab:hyst}. To determine the shape and size of the NdFeB particles, Scanning Electron Microscope (SEM) images were made, see Fig.\ref{fig:neofer_sem}. The particles are of spherical morphology with a diameter of approximately 50$\pm$20\,\textmu m.
%The datasheet of the compound exhibits a remanence $B_r$ of 400\,mT, intrinsic coercivity $H_{cj}$ of 630\,KA/m, and a  maximum energy product $(BH)_{max}$ of 27\,KJ/m$^3$.

For the 3D printer setup the Neofer\,\textregistered$ $ 25/60p compound granules with a size of around 5\,mm have to be extruded to filaments with a diameter of 1.75$\pm$0.1\,mm. The extrusion is performed with a Leistritz ZSE 18 HPe-48D twin-screw extruder, and a Sikora Laser Series 2000 diameter measuring system to control the diameter tolerances of the filament.

\section{Printer}
An important aim of this work is to manufacture isotropic NdFeB permanent magnets with high dimensional accuracy. We use a commercially available fused deposition modeling printer. This system creates the object layer by layer by a meltable thermoplastic. The wire-shaped plastic is first heated to just above its softening point. With the aid of an extruder, and a movable, heated nozzle the object is built up in layers on the already solidified material on the building platform \cite{3d_paper, practical_3d_printers}.

We choose the Builder 3D printer from Code P. The maximum building size is 220$\times$210$\times$164\,mm (L$\times$W$\times$H), the layer height resolution is 0.05\,-\,0.3\,mm. The nozzle diameter is 0.4\,mm, and it is fed with filaments with a diameter of 1.75\,mm. To control the printer, the open-source software Repetier-Host and Slic3r are used. For the optimal printing results with the NdFeB compound material, which is described above, the best empirically found printing and slicer parameters are listed in Tab. \ref{tab:print_parameter}. 

\begin{table}[htbp]
\centering
\begin{tabular}{m{2.8cm}|m{4cm}}
\textbf{Parameter} & \textbf{Value}\\\hline
Extruder temp. & 255\,$^\circ$C \\
Layer height & 0.1\,mm \\
Printer speed & 20\,mm/s \\
Fill density & 100\,\% \\
Fill pattern & Rectilinear \\
Bed adhesion & Solvent free Pritt glue stick \\
Bed temp. & 40\,$^\circ$C \\
\end{tabular}
\caption{Best empirically found printer and slicer parameter for Neofer\,\textregistered$ $ 25/60p.}
\label{tab:print_parameter}
\end{table}

\section{3D Magnetic Field Scanner}
To measure the stray field of the printed permanent magnet, the 3D printer is upgraded to a 3D magnetic flux density measurement system. As sensing device, a 3D Hall sensor TLV493D-A186 from Infineon is used.  The microcontroller is programmed to read out the components of $\vec{B}$ with a frequency of 0.5\,kHz. The sensor has a measurement range of $\pm$130\,mT, and a measured detectivity of 40\,\textmu T/$\sqrt{Hz}$ for DC magnetic fields.
%with a resolution of 97\,\textmu T. However, with oversampling the resolution of the sensor can be increased. In our system an averaging over 512 measurements leads to a resolution of 5\,\textmu T with a measuring frequency of 1\,Hz. 
%For an easy implementation, the evaluation board 3D Magnetic 2Go with an integrated XMC4200 microcontroller and USB connection are chosen.
The adjustment and the alignment of the sensor is important for the accuracy of the stray field measurement. The idea is to avoid an exact positioning and alignment of the sensor, because this would lead to a complex sensor mounting system. A calibration method relying on detailed stray field simulations is proposed. In our case the sensor is attached to the extruder head with a self printed suspension without any adjustment.

The calculated magnetic flux density $\vec{B}$ of a predefined object (e.g. cylinder magnet) is used to calibrate the sensor. The main idea of the calibration is to use an inhomogeneous reference field which relates to the geometry of the calibration sample and its stray field. The measurement system scans the magnetic field in a defined volume outside the cylinder magnet. With the numerical solution $\vec{B}_{s}$ and the measured field $\vec{B}_{m}$ an optimization problem can be formulated, where the squared error between the calculated and measured values is minimized, in order to determine the calibration parameters.

A reference field of a cylinder magnet can be calculated numerically by solving $\vec{B}_{s}(\vec{r}) = -\mu_0 \nabla\Phi$ with the magnetic potential $\Phi(\vec{r})$ \cite{jackson}.Where $\mu_0$ and $\vec{M}$ are the vacuum permeability and the magnetization, respectively. The magnetization is chosen along the z-axis ($\vec{M}=M_0\cdot\vec{e}_z$).
%\begin{align}
%\Phi(\vec{r}) &= -\frac{1}{4\pi} \int \frac{ \nabla \cdot \vec{M}}{| \vec{r}-\vec{r}_i |} \textrm{d}\vec{r}_i \\
% \vec{B}_{s}(\vec{r}) &= -\mu_0 \nabla\Phi
%\end{align}

The experimental field $\vec{B}_{exp}$ can be calculated with the measurement data $\vec{B}_m(\vec{r})$ and the background field $\vec{B}_{back}(\vec{r})$ to $\vec{B}_{exp} = \vec{B}_m(\vec{r}) - \vec{B}_{back}(\vec{r})$. $\vec{B}_{back}(\vec{r})$ is a baseline measurement to reduce the systematic errors of the setup.
%\begin{align}
% \vec{B}_{exp} &= \vec{B}_m(\vec{r}) - \vec{B}_{back}(\vec{r})  .
%\end{align}
The simulated field $\vec{B}_{sim}$ in relation to the calibration parameters $\vec{s},\gamma,\beta,\alpha,\vec{\Delta r}$ is defined as:
\begin{align} 
 \vec{B}_{sim} &= \vec{s}\cdot \textbf{R}_{ZYX}(\gamma,\beta,\alpha)\cdot\vec{B}_{s}(\vec{r}-\vec{\Delta r})
\end{align}
where $\vec{s}$ is the sensitivity of the Hall sensor, $\textbf{R}_{ZYX}$ is the rotation matrix with the Euler angles $(\gamma,\beta,\alpha)$ to compensate the tilting of the sensor, and $\vec{\Delta r}$ is the sensor offset.

Solving the following minimization problem results in the unknown parameters of the calibration:
\begin{align}
 \min_{\vec{s},\gamma,\beta,\alpha,\vec{\Delta r}}\Arrowvert \vec{B}_{exp} - \vec{B}_{sim} \Arrowvert^2  .
\end{align}
Powell's method is used for minimization \cite{powell}.% Tab. \ref{tab:calib} lists the calibration parameters for our setup.
%The algorithm is simple to implement in Python with the Scipy package. The function \textit{``scipy.optimize.fmin\_powell''} is a modification of Powell's method to find the minimum of a function of $N$ variables. Powell’s method is a conjugate direction method \cite{powell}.
% \begin{table}[htbp]
% \centering
% \begin{tabular}{m{1.9cm}|m{3cm}}
% \textbf{Parameter} & \textbf{Value}\\\hline
% $\vec{s}$ (\,)& 0.97/0.95/0.76 \\
% $\gamma,\beta,\alpha$ ($^\circ$)& 1.9/0.9/2.3\ \\
% $\vec{\Delta r}$ (mm)& 0.427/0.010/-0.108
% \end{tabular}
% \caption{Calculated calibration parameters for the 3D field scanner.}
% \label{tab:calib}
% \end{table}

\begin{figure}
	\centering
	\includegraphics[width=\linewidth]{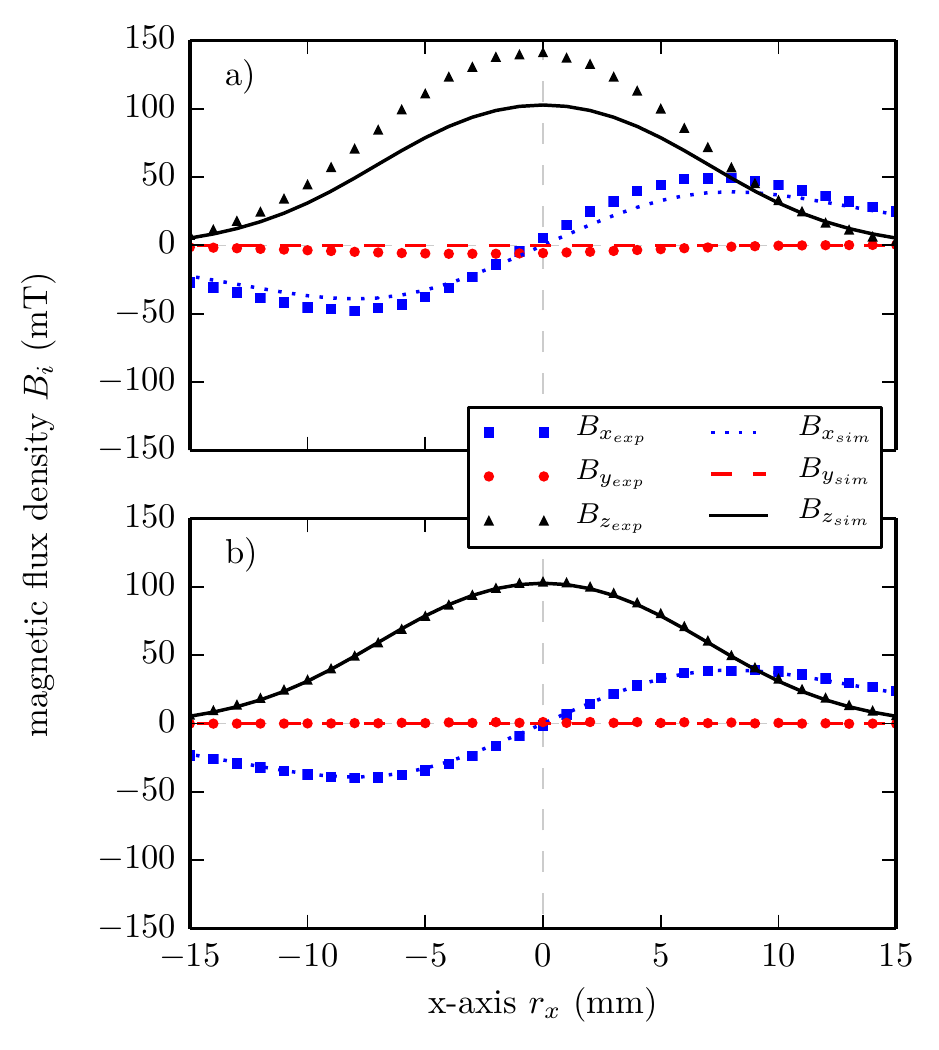}
	\caption{(a) Line scan ($y=0$, $z=16$\,mm) of $\vec{B}_{exp}$ above the cylinder magnet ($d=15$, $l=8$\,mm) without calibration, and the simulation $\vec{B}_{sim}$ at the same line. (b) Measurement data in comparison with the simulation after calibration.}
	\label{fig:calib_comp}
\end{figure}

Fig. \ref{fig:calib_comp} a) indicates a significant difference between simulation and measurement. After applying the calibration procedure (b), the measurement fits very well with the simulation after the calibration method. The 3D Hall sensor has a X/Y to Z sensitivity deviation of approximately 20\,\%.

%The sensor movements, and therefore the measuring points in the space, are controlled with a Python script, as well as to read out and save the sensor data.
With this method the field $\vec{B}$ can be scanned in 1D, 2D, and even 3D outside an object. The spatial resolution of the printer is 0.1\,mm for the x and y-axis, and 0.05\,mm for the z-xis.  

\section{Results}
\begin{figure}
	\centering
	\includegraphics[width=\linewidth]{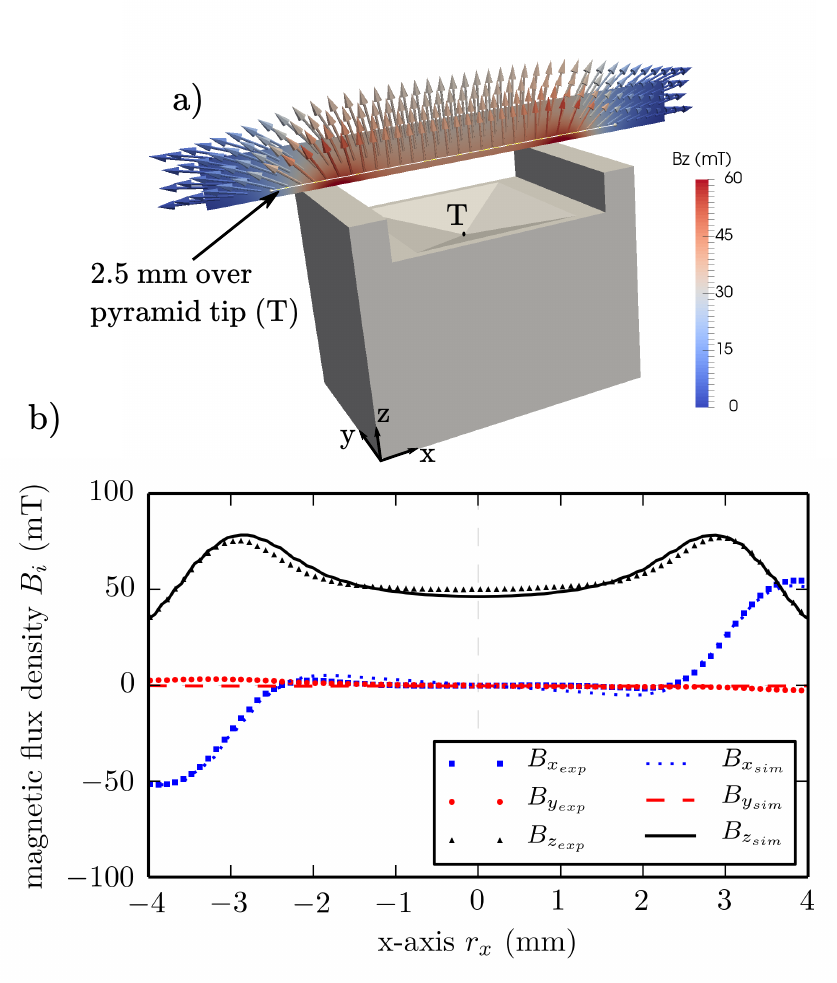}
	\caption{(a) Geometry of the permanent magnet, and area scan of $\vec{B}$ with a step size of 0.1\,mm in the middle of the printed magnet. (b) Line scan 2.5\,mm over pyramid tip (T) compared with FEMME simulation of the magnet with perfect shape.}
	\label{fig:bp_mag}
\end{figure}

The printing process is benchmarked with a complex geometry that is known to minimize the components of the magnetic stray field $\vec{B}$ in x and y direction in a wide range along the x-axis $r_x$. This is an important aspect for GMR sensor applications \cite{ibb}. The printed magnet is magnetized along the z-axis inside a pulse coil with 4\,T. Finite element simulations using FEMME were performed to find the best design for the magnet \cite{femme, femme2}. The numerical simulation solving the magnetostatic Maxwell equations, and using a vector hysteresis model for a proper description of the permanent magnetic material. A rectangular magnet with a pyramid notch shows the best simulation results (Fig. \ref{fig:bp_mag} a).

\begin{figure}
	\centering
	\includegraphics[width=0.76\linewidth]{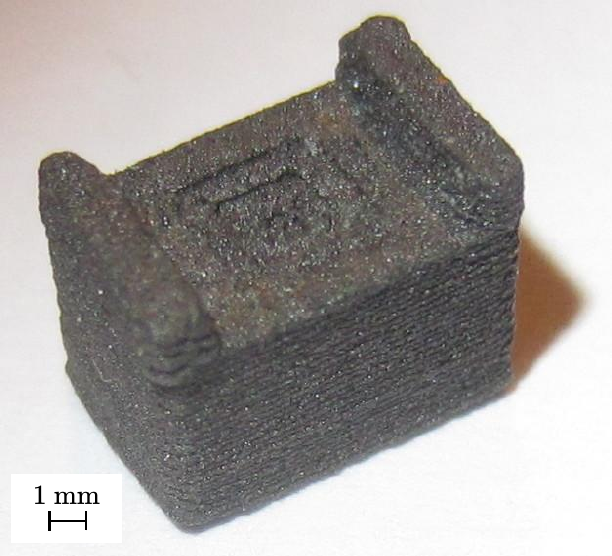}
	\caption{Printed isotropic NdFeB magnet with Neofer\,\textregistered$ $ 25/60p with optimized shape to suppress $B_x$ and $B_y$ along the x-axis $r_x$.}
	\label{fig:bp_print}
\end{figure}

Fig. \ref{fig:bp_print} shows a picture of the printed magnet with the isotropic Neofer\,\textregistered$ $ 25/60p material. The overall size of the magnet is 7$\times$5$\times$5.5\,mm (L$\times$W$\times$H) with a layer height of 0.1\,mm, and features with a thickness of 0.8\,mm. This indicates the possibility to print miniaturized magnets with complex structures. Fig. \ref{fig:bp_mag} a) displays an area scan of $\vec{B}$, and a line scan above the pyramid tip (T). Compared with the FEMME simulation, it points out a good conformity between printed and simulated magnet.

\begin{figure}
	\centering
	\includegraphics[width=1\linewidth]{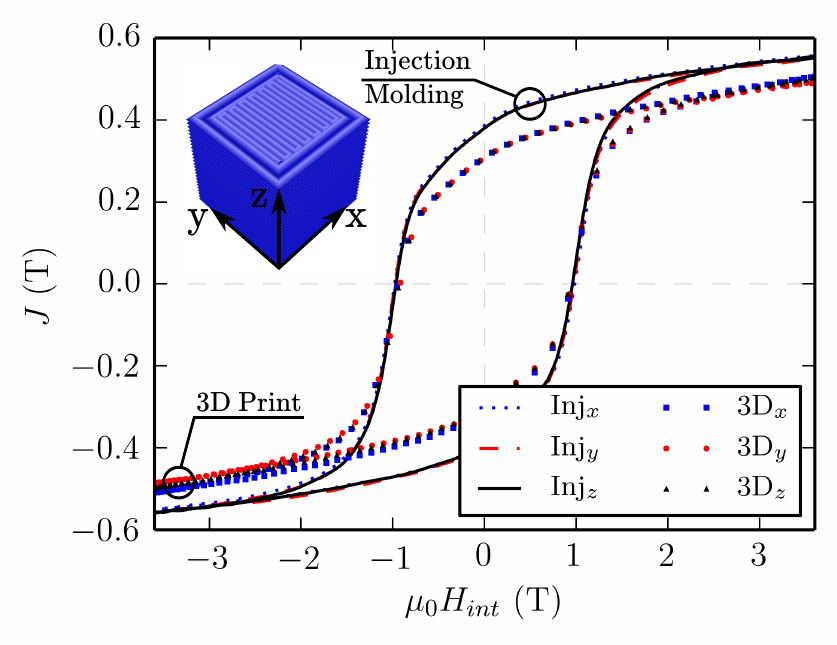}
	\caption{Hysteresis measurement of injection molded and 3D printed cube in different magnetization directions.}
	\label{fig:hysterese}
\end{figure}

Magnetic properties of Neofer\,\textregistered$ $ 25/60p polymeric magnetic composite materials were measured by Pulsed Field Magnetometry (PFM) (Hirst PFM11) \cite{pfm2, pfm}. All measurements were carried out with the same parameters - temperature of 297\,K and a magnetic field up to 4\,T peak field. Cubes with an edge length $a$ of 5$\pm$0.02\,mm are prepared. The hysteresis measurements are performed in x, y, and in z magnetization direction to identify the magnetic behaviors of the layer structure (see Fig. \ref{fig:hysterese}) of the printed magnet. The average demagnetisation factor $N$ for a cube is 1/3 \cite{demag}. Fig. \ref{fig:hysterese} shows the hysteresis measurement of the injection molded and the 3D printed cube for each magnetization direction. The injection molded sample shows an isotropic behavior, like the 3D printed cube which indicates, that the layer structure of the printing process is irrelevant for the magnetic properties. The measured remanence $B_r$, the intrinsic coercivity $H_{cj}$, and the volumetric mass density $\varrho$ of the samples are listed in Tab. \ref{tab:hyst}. $B_r$ differ between the datasheet and the measured value around 4\,\%. The deviation of the volumetric mass density $\varrho$ between the injection molded and the printed cube is 22\,\% and therefore in the same range as the deviation of $B_r$ of 25\,\%.

\begin{table}
\centering
\begin{tabular}{m{2cm}|m{1.8cm} m{1.8cm} m{1.8cm}}
\textbf{Procedure} & $\pmb{\varrho}$ (g/cm$^3$) & $\pmb{B_r}$ (mT) & $\pmb{H_{cj}}$ (kA/m) \\\hline
datasheet & 4.35 & 400 & 630\\
3D print & 3.57 & 310 & 740\\
inj. mold. & 4.35 & 387 & 771
\end{tabular}
\caption{Summery of the datasheet and the measured Neofer\,\textregistered$ $ 25/60p material properties.}
\label{tab:hyst}
\end{table}

\section{Conclusion \& Outlook}
A novel method to manufacture polymer bonded isotropic magnets is presented with no special instruments or highly expensive equipment is necessary for this process. A 3D printer is used and upgraded to a full 3D field scanner. With our calibration method, an elaborate adjustment is no longer required. With a simulation the calibration parameters are easy to calculate. The effectiveness of the method is presented. 

As an example we present results of a printed magnet with special magnetic properties. For application for wheel speed sensing permanent magnets are required that produce a strong field parallel to the magnetization ($B_z$) but a fields as small as possible in orthogonal direction ($B_x$) along a line in x-direction above the magnet. 

The performance of the printed magnet is evaluated by scanning the  field of the printed magnet and comparing it with finite element simulations solving the macroscopic Maxwell equations. Excellent agreement could be obtained indicating that indeed the magnet with the desired shape and magnetic properties could be printed. In addition hysteresis measurements of a printed magnet were compared with injection molded samples. The lower volumetric mass density of the printed magnet leads to a lower remanence of the printed polymer bonded magnet. Detailed studies how the  volumetric mass density can be controlled by the printer parameter like overlap or line width, is subject to future research. 

The presented fabrication method using dual extruder can be used to print magnets  composed of locally different polymer matrix materials, as well as different magnetic powder ranging from soft magnetic alloys to hard magnetic NdFeB or ferrite alloys. The ability to print magnets with locally varying magnetic materials with tailored magnetic properties opens a new door for applications with field profiles and magnetic properties which can not be produced with state of the art methods.

\section{Acknowledgement}
The support from CD-laboratory AMSEN (financed by the Austrian Federal Ministry of Economy, Family and Youth, the National Foundation for Research, Technology and Development) is acknowledged. The authors would like to thank Magnetfabrik Bonn GmbH for the provision of the compound material, and to Montanuniversitaet Leoben and the EU Project Horizon 2020 REProMag (grant No. 636881). The SEM and sample preparations are carried out using facilities at the University Service Centre for Transmission Electron Microscopy, Vienna University of Technology, Austria.

\newpage

\bibliographystyle{aipnum4-1}
%\bibliography{pub16_1}


\end{document}